\documentclass[prl,twocolumn,amsmath,amssymb,showpacs]{revtex4}
\usepackage{graphicx}
 
\begin{document}
 
\title{Comment on ``Nonmagnetic Impurity Resonances as a Signature of 
Sign-Reversal Pairing in Fe-As-Based Superconductors''}

\author{Maria Daghofer}
\email{M.Daghofer@ifw-dresden.de} 
\affiliation{IFW Dresden, P.O. Box 27 01 16, D-01171 Dresden, Germany}

\author{Adriana Moreo}
 
\affiliation{Department of Physics and Astronomy,University of Tennessee,
Knoxville, TN 37966-1200 and
\\Oak Ridge National Laboratory,Oak Ridge,
TN 37831-6032,USA}

\maketitle

In a recent Letter\cite{zhang}, the energy band structure of Fe-As-based 
superconductors is fitted with a tight-binding model with two Fe ions per 
unit cell and two degenerate $d_{xz}$ and $d_{yz}$ orbitals  per Fe ion. 
The author claims that the proposed model, which differs markedly from
a model previously used by other authors for the same two orbitals in
the same compounds~\cite{scalapino, ours}, possesses the symmetry
required to describe the Fe-As planes in iron-pnictide
superconductors. In this comment we argue that this is not the case. 

As discussed in Ref.~\onlinecite{zhang}, the unit cell of the Fe-As
planes contains two iron ions, and the Hamiltonian Eq.~(1) employed
there reflects this. However, each two-iron unit cell also has an
internal symmetry that the Hamiltonian needs to obey. To see this, 
consider the square lattice made up by all iron ions, i.e.,
with one iron per unit cell, and the Fe-Fe distance as a new basis
vector. The Fe-As planes are invariant under (a) translation by one
unit along the Fe-Fe direction, followed by (b) a reflexion on the
Fe-Fe plane. At first sight, the Hamiltonian of
Ref.~\onlinecite{zhang} appears to take this into account, because a
translation by one Fe-Fe distance and an additional exchange of
hoppings $t_2$ (mediated by an As \emph{above} the plane) and $t_3$ (via an
As \emph{below} the plane) indeed leaves it invariant. But since the
Fe-As-Fe distances and angles are the same for As ions above and below
the plane, these two hopping paths are very symmetric, which induces
additional restrictions~\cite{eschrig} and the
two paths actually give the same hopping
~\cite{note_t4}. 

This missing symmetry of the Hamiltonian also reveals itself in the
momentum-dependent band structure. The $d_{xz}$ and $d_{yz}$ orbitals
should be degenerate at the $\Gamma$ 
point~\cite{moreo,eschrig}, as it can also be seen in band structures
obtained from the local density approximation~\cite{first,singh,xu,cao,fang2} 
or by a Slater-Koster approach~\cite{moreo,calderon}. Another 
consequence of the internal symmetry of the two-iron unit cell is that
all bands have to be two-fold degenerate at the boundaries of the
Brillouin zone (BZ) corresponding to the two-Fe unit cell~\cite{eschrig}, which
is clearly violated in the Hamiltonian presented in
Ref.~\onlinecite{zhang}, see the data for the $(\pi,0)$-$(\pi,\pi)$
path in Fig.~(2).\cite{foot} In
fact, any tight-binding Hamiltonian for the Fe-As planes can be 
written in block form, where each block is expressed in terms of the 
orbitals of 
\emph{one single Fe per unit cell}~\cite{moreo,eschrig}. Instead of studying 
both blocks
for the BZ of the original two-iron unit cell, one can
then consider just one block, but in the extended BZ
corresponding to a unit cell with just one Fe
ion, because the two blocks correspond to momenta ${\bf k}$ and ${\bf
  k}+(\pi,\pi)$ in the extended BZ.  In this description, the system has
$D_{4h}$  
symmetry which means that all the eigenstates at the center of the BZ should
transform according to irreducible representations of $D_{4h}$. In particular
the orbitals $d_{xz}$ and $d_{yz}$ transform according to the two dimensional 
representation $E_{g}$ leading to the degeneracy at the center of the
BZ, which is missing in the Hamiltonian of Ref.~\onlinecite{zhang}.

Finally, $t_3$ is five times larger than $t_2$ in
Ref.~\onlinecite{zhang}. This difference is extreme even for the
orthorhombic phase of the pnictides, where a slight oxygen distortion
of 1\% (6\%) in bulk (on surfaces)~\cite{vonbraun} might in principle
justify a small difference.   

This work was supported by the NSF grant DMR-0706020 and the
Division of Materials Science and Engineering, U.S. DOE, under contract
with UT-Battelle, LLC.


\begin{thebibliography}{}
\bibitem{note_t4}
Additionally, the sign of $t_4$ should depend on the hopping
direction. It may be worth noting that going to a different orbital
basis does not restore the symmetry of the Hamiltonian, as can be seen
from the discussion of the momentum-dependent bands, which
do not depend on the orbital basis.
\bibitem{foot} This applies to 122 compounds when a  
two-dimensional description is used.\cite{eschrig}
\end{thebibliography}
\end{document}